\documentclass[twocolumn,prd,floatfix]{revtex4}

\usepackage{physics}
\usepackage{color}
\usepackage{tabularx}
\usepackage{epsfig}
\usepackage{amsmath}
\usepackage{amssymb}
\usepackage{bm}
\usepackage{graphicx}
\usepackage{multirow}

\usepackage{epsfig}

\begin{document}
	\title{Quantum Steering Ellipsoid and Unruh Effect}

	\author{ Yusef Maleki,$^1$ Bahram Ahansaz,$^2$ Kangle Li,$^3$  Alireza Maleki$^4$
		\\{\small$^1$ Department of Physics and Astronomy,
			Texas A\&M University, USA}
		\\{\small$^2$ Physics Department, Azarbaijan Shahid Madani University, Tabriz, Iran}
		\\{\small$^3$ Department of Physics, Hong Kong University of Science and Technology, Hong Kong SAR, China
		}
			\\{\small$^4$ Department of Physics, Sharif University of Technology,
			Tehran, Iran}
	}

	\date{\today}

	\begin{abstract}
Quantum steering is a perplexing feature at the heart of quantum mechanics that provides profound implications in understanding the nature of physical reality. On the other hand, the effect of relativistic features on quantum systems is vital in understanding the underlying foundations of physics. In this work, we study the effects of Unruh acceleration on the quantum steering of a two-qubit system. In particular, we consider 
the so-called quantum steering ellipsoid and the maximally-steered coherence in a non-inertial frame
and find closed-form analytic expressions for the role of the Unruh acceleration in these quantities.
Analyzing the conditions for the steerability of the system, we develop a geometric description for the effect of Unruh acceleration on the quantum steering of a two-qubit system.

	\end{abstract}

	\maketitle
	
	\newpage

	\section{Introduction}
		Quantum physics is one of the most important scientific achievements in the history of science which is based upon bizarre and counterintuitive features. One of these bizarre features was first addressed by Einstein, Podolsky, and Rosen in 1935, which is known as the EPR paradox \cite{einstein1935can}. The criticism of the EPR paradox resulted in remarkable progress in unraveling various quantum mechanics features. In the same year, analyzing the EPR paradox,  Schr\"odinger introduced the concept of quantum steering \cite{schrodinger1935discussion,uola2020quantum}. To illustrate the concept of quantum steering, one can consider two physicists such as  Alice and Bob, who share a bipartite two-qubit system. In other words, Alice has one of the qubits in her possession while Bob has the other qubit in his possession. Now, when Alice makes a measurement with several possible outcomes, then Bob will be left with a conditional reduced density matrix depending on what outcome Alice obtains. Alice can choose a different basis for the measurement of her subsystem, which affects the density matrix that Bob's qubit reduces to.
	This process is called the "steering". 
	From a geometric perspective, a single qubit state can be represented by a vector on a Bloch ball.
	Considering the quantum steering protocol, the state of the steered qubit can be represented as a vector in the unit Bloch ball. Therefore, the steering process provides the region inside the unit Bloch ball that the steered state can be represented. The region has been found to be an ellipsoid, and it is called the "Quantum Steering Ellipsoid" \cite{verstraete2002quantum_qse, shi2011quantum_qse, jevtic2014qse, jevtic2015einstein_qse}.  
	 
 Quantum steering ellipsoid (QSE) is geometrically a natural generalization of Bloch ball, and it provides new insights into the quantum entanglements and steering. Quantum steering has been studied extensively in recent years, and it exhibits remarkable potential in understanding different entanglement structures and has important implications in quantum communications and cryptography \cite{branciard2012one,cavalcanti2015detection,acin2007device}.

On the other hand, considering the fundamental laws of nature, it appears that in the regions where the effects of both quantum mechanics and general relativity become significant, a reliable description of the systems encounters some dramatic challenges.
	 For example, studying the quantum effects near the event horizon of black holes results in controversial concepts such as the information paradox \cite{penington2020entanglement,almheiri2019entropy,almheiri2020page}. To settle such issues, a comprehensive understanding of the behavior of quantum description of the systems in the presence of a strong relativistic effect is required \cite{hawking1976breakdown,susskind1997black,mathur2009information,scully2018quantum}.

	Since the event horizon of the black hole and generally any nonzero curvature can be approximated with the locally accelerated reference frame, the study of quantum mechanical effects in the non-inertial references frames is of significant importance. The study of quantum mechanics in the accelerated reference frame was first introduced by Unruh \cite{unruh1976notes}, where he showed that observers in the non-inertial reference frames experience the vacuum as a thermal bath associated with the temperature known as the Unruh temperature. 
 This concept is quite similar to the description of the black hole physics leading to Hawking temperature and consequently black hole evaporation. It is notable that the experimental detection of the Unruh effect is a long-lasting question and has critical importance in the study of quantum gravity \cite{Unruh1981}.

	  In recent years, there has been a considerable effort for observing the Unruh effect in various settings, ranging from accelerated quantum systems \cite{lima2019probing,wang2021coherently} to the study of analog gravitational systems in the condensed matter physics \cite{unruh1995sonic,unruh2008dumb,holes2002m}. Therefore, the investigations of the systems such as acoustic black holes \cite{Unruh1981},  and Bose-Einstein condensate analog of black holes \cite{garay2000sonic,lahav2010realization} has offered new insights into the nature of the non-inertial effects on the quantum system \cite{belgiorno2010hawking,steinhauer2014observation,steinhauer2016observation,boiron2015quantum, visser1998acoustic}.
	  On the other hand, there has been significant progress in observing the physics of black holes,  from signals of strong gravitational binary system merging in the LIGO  gravitational wave detection \cite{abbott2016observation} to the first image of the black hole in the Event Horizon Telescope (EHT) \cite{event2019first}, making the extraction of observational data from the strong gravitational system such as black holes practical.

	In this work, we consider the entanglement and the so-called quantum steering ellipsoid and the maximally-steered coherence in a non-inertial frame. We map the steered state into Bloch ball and find the relation between the shape of the ellipsoid and the reference frame acceleration.
	 More specifically, we quantify the impact of the Unruh effect on the quantum steering and the maximally-steered coherence of the bipartite quantum states.	Considering the entanglement of the system, we show that entanglement reduces in the accelerated frame.

	\section{Unruh effect for  fermionic single qubit}

	We first introduce the fermionic field in an inertial frame that we call laboratory frame, which is characterized by the coordinate parameters $(t,x)$. In the Minkowskian 3+1 spacetime the free field for a fermion obeys the Dirac equation \cite{schwartz2014quantum}
	
	\begin{align}
	i\gamma^{\mu}\partial_{\mu}\psi-m\psi=0,
	\label{Dirac}
	\end{align}
	
	where $m$ is the mass of the particle and $\gamma^{\mu}$ is the four gamma matrices. This field is quantized on the basis of the positive and negative mode, which we respectively represent by $\psi_{\textbf{k}}^+$ and $\psi_{\textbf{k}}^-$, in which $\textbf{k}$ denotes the wavenumber. Hence, with this definition, the basis of the field could be written as \cite{AlsingDF2006}
	
	\begin{align}
	\psi_{\textbf{k}s}^\pm=\frac{1}{\sqrt{2\pi \omega_{\textbf{k}}}}\phi^\pm_{\textbf{k}s} e^{\pm i(\textbf{k.x}-\omega_{\textbf{k}}t)},
	\label{basispm}
	\end{align}
	where $\omega_{\textbf{k}}=\sqrt{m^2+\textbf{\textbf{k}}^2}$, and $\phi_{\textbf{k}s}$ indicates the spinor amplitude. Also, $s$,  which labels the spin state, could be either up or down $(\uparrow,\downarrow)$.

	Therefore, we can express the Dirac field in this basis as
	\begin{align}
	\psi=\int d\textbf{k}(a_{\textbf{k}}\psi_{\textbf{k}}^+ +b_{\textbf{k}}^\dagger \psi_{\textbf{k}}^-).
	\label{qDirac}
	\end{align}
	where, the operators $a_{\textbf{k}}^\dagger,b_{\textbf{k}}^\dagger$ and $a_{\textbf{k}},b_{\textbf{k}}$ are the creation and the annihilation operators for the particle and the anti-particle solution of Dirac equation for momentum $\textbf{k}$ respectively. These operators satisfying the fermionic anticonmutaion relations
	\begin{align}
	\lbrace a_{i},a_{j}^\dagger\rbrace=\lbrace b_{i},b_{j}^\dagger\rbrace=\delta_{ij},
	\label{Manticonmutaion}
	\end{align}
	and all other anticommutators vanish. 
	
	The  vacuum state in the Minkowski framework is defined as the state with no excitations in an inertial frame
	
	\begin{align}
	\vert 0\rangle=\prod_{\textbf{k}\textbf{k}'} \vert 0_{\textbf{k}}\rangle^+ \vert 0_{\textbf{k}}\rangle^-,
	\label{vacuum}
	\end{align}
	where $ a_{\textbf{k}}\vert 0_{\textbf{k}}\rangle^+= b_{\textbf{k}}\vert 0_{\textbf{k}}\rangle^-=0$, defines the vacuum state for each mode $\textbf{k}$ and $ (a_{\textbf{k}}^\dagger)^2= (b_{\textbf{k}}^\dagger)^2=0$, indicates that only two particles can be created in each mode.
	
	Now, we will consider this field being observed concerning a non-inertial observer moving with the proper acceleration $a$. To describe the phenomena in this accelerated reference frame, we need to introduce the proper coordinates $(\tau,\xi$), in which $\tau$ is called proper time and $\xi$ is the distance measured by the accelerated observer. These coordinates are called Rindler coordinates, and both range from $-\infty$ to $+\infty$. In order to find the relation between the two coordinate systems, we find   the trajectory of an accelerated observer with respect to the inertial laboratory  frame as \cite{CarrollBOOK,mukhanov2007introduction}
	
	\begin{align}
	x=\frac{e^{a\xi}}{a}\sinh(a\tau), \quad t=\frac{e^{a\xi}}{a}\cosh(a\tau).
	\label{M-R}
	\end{align}
	
\begin{figure}
		\centering
		\includegraphics[width=3.5 in]{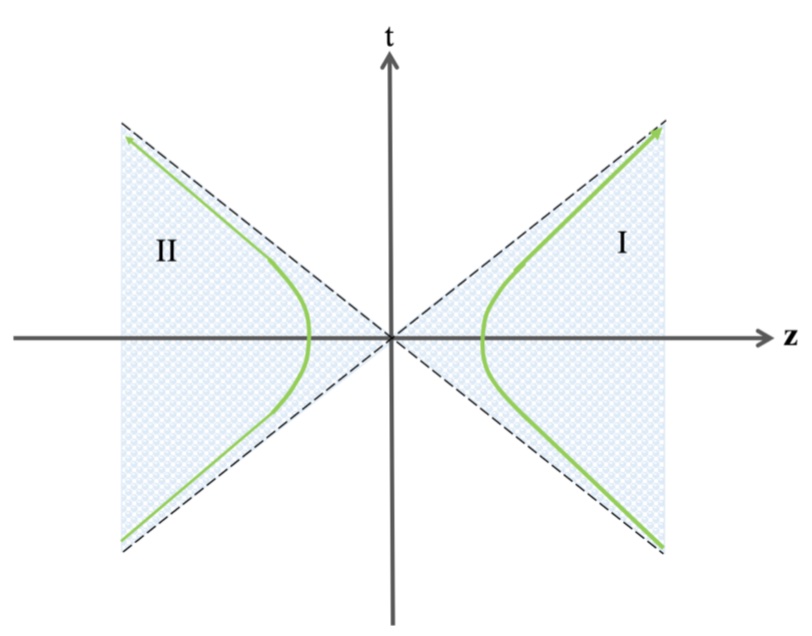}
		\caption{ Rindler space-time diagram with one spatial dimension. Here the curve is plotted for a particle with a constant acceleration. The two regions I and II are causally disconnected.
		}
		\label{fig:1}
	\end{figure}

	Therefore, the accelerated observer's equation of motion is a hyperbola ($x^2-t^2=\frac{e^{a\xi}}{a}$) for the line with constant $\xi$. Furthermore, from the fraction $\frac{x}{t}=\coth(a\tau)$, we find that the constant $\tau$ is a straight line passing through the origin, and for $\tau\longrightarrow \pm \infty$ the observer's velocity reaches the speed of light, and the slope goes to $\pm1$. These are depicted in Fig. \eqref{fig:1}. Hence, the accelerated observer confined in the region between the two lines determines the Rindler wedge ($I$). This region covers only a quarter of the Minkowski spacetime. We can also define the second set of coordinates for the causally disconnected region (Rindler wedge ($II$)) as \cite{CarrollBOOK,mukhanov2007introduction}
	
	\begin{align}
	x=-\frac{e^{a\xi}}{a}\sinh(a\tau), \quad t=-\frac{e^{a\xi}}{a}\cosh(a\tau).
	\label{RindlerXT}
	\end{align}
 These two regions are causally disconnected by two lines with the slope of one that is called Rindler horizon. Thus,  we have two separate bases for regions $I$ and $II$, that we could represent by $\lbrace \psi_{\textbf{k}}^{I+},\psi_{\textbf{k}}^{I-}\rbrace$,$\lbrace \psi_{\textbf{k}}^{II+},\psi_{\textbf{k}}^{II-}\rbrace$, respectively. These two bases form a complete basis for describing the Dirac field \eqref{Dirac} in the Rindler spacetime \cite{CarrollBOOK,mukhanov2007introduction}
	
	\begin{align}
	\psi=\int d\textbf{k}(c_{\textbf{k}}^{I+}\psi_{\textbf{k}}^{I+} +d_{\textbf{k}}^{I\dagger}\psi_{\textbf{k}}^{I-}+c_{\textbf{k}}^{II}\psi_{\textbf{k}}^{II+}+d_{\textbf{k}}^{II\dagger}\psi_{\textbf{k}}^{II-}),
	\label{RDirac}
	\end{align}
	where $(c_{\textbf{k}}^{\sigma},c_{\textbf{k}}^{\sigma\dagger})$ are the creation and the annihilation operators for the fermionic particles and $(d_{\textbf{k}}^{\sigma},d_{\textbf{k}}^{\sigma\dagger})$ are the creation and the annihilation operators of the antiparticle fermions. Here, $\sigma$ denotes  the Rindler wedge with $\sigma \in \lbrace I,II \rbrace$. These operators obey the  anticommutation  relation similar to \eqref{Manticonmutaion}
	
	\begin{align}
	\lbrace c_{\textbf{k}}^{\sigma},c_{\textbf{k}}^{\sigma'\dagger} \rbrace=\lbrace d_{\textbf{k}}^{\sigma},d_{\textbf{k}}^{\sigma'\dagger} \rbrace=\delta_{\sigma'\sigma}\delta_{\textbf{k}\textbf{k}'}
	\label{Ranticonmutaion}
	\end{align}
	The relation between the creation and annihilation operator of the Dirac filed in Minkowsi, and Rindler coordinate can be found in \cite{AlsingDF2006}, that is called Bogoliubov transformation. This transformation, for example,  gives the $a_{\textbf{k}}$ as a linear combination of $ c_{\textbf{k}}^{I}$ and $d_{\textbf{k}}^{II\dagger}$ with some coefficients. We use the single-mode approximation in which we suppose that the accelerated observe carries a detector that is sensitive just to the same wavenumber $\textbf{k}$, that is observed by the lab observer \cite{bruschi2010unruh}. Thus, we shall not repeat the $\textbf{k}$ label anymore in the rest of the paper. One can construct, using this transformation, the Minkowski vacuum in terms of the Rindler basis for a  positive mode (particles) with wavenumber $\textbf{k}$ as \cite{AlsingDF2006,bruschi2010unruh}

	\begin{align}
	\vert 0\rangle_{M}= \cos(r)\vert 0\rangle_{I} \vert 0\rangle_{II}+\sin(r)\vert 1\rangle_{I} \vert 1\rangle_{II},
	\label{MR0}
	\end{align}
	where $r=\exp(\frac{\pi \omega c}{a})$. Also, by acting $a_{\textbf{k}}^{\dagger}$ on this state we can  expressed the exited state as
	\begin{align}
	\vert 1\rangle_{M}= \vert 1\rangle_{I} \vert 0\rangle_{II}.
	\label{MR1}
	\end{align}
	Therefore, we use the tensor product of two sets of Fock states to express the states given in Minkowski coordinate in the Rindler coordinate.

	\section{ Unruh effect and Quantum correlations in two-qubit systems}

	Now, we are going to consider quantum steering ellipsoid of two-qubit states in an accelerating reference frame.   
	To this end, let us consider Alice and Bob share a  two-qubit Werner state given by \cite{werner1989quantum,miranowicz2004violation}
	\begin{align}
	\rho=\frac{(1-p)}{4}I+p |\psi \rangle \langle \psi|.
	\end{align}
	Here, the state $|\psi \rangle$ is a two-qubit maximally entangled state
	\begin{align}
	|\psi \rangle=\frac{1}{\sqrt{2}}(|00 \rangle+|11 \rangle).
	\end{align}
	 This class of states was first constructed by Werner in 1989 to show that it is entangled but not Bell nonlocal for $1/3<p<1/2$. 
	The matrix representation of this density matrix in the logical bases is given by
	
	\begin{eqnarray}\label{wl-rho}
	\rho=\frac{1}{2}\left(
	\begin{array}{cccc}
	\displaystyle \frac{1+p}{2} & 0 & 0 & \displaystyle p\\
	0 & \displaystyle \frac{1-p}{2} & 0 & 0\\
	0 & 0 & \displaystyle \frac{1-p}{2} & 0\\
	\displaystyle p  & 0 & 0 & \displaystyle \frac{1+p}{2}\\
	\end{array}
	\right).
	\end{eqnarray}
	In this setting, the parameter $p$ could vary from 0 to 1, where for $p=0$ we attain the maximally mixed two-qubit state that is a $4\times 4$ unit matrix with no entanglement. For $p=1$ we attain the maximally entangled state $|\psi \rangle$. Note that this state is entangled if and only if  $p>1/3$. Therefore, this class of states can present two-qubit states with the entanglement ranging from zero to maximum. Thus, providing an interesting set of states for exploring quantum correlations.

	Now, let us consider the scenario such that the first qubit belongs to Bob and the second qubit to Alice. Furthermore, we assume that Bob is uniformly accelerated with the constant acceleration $a$. Therefore, the density matrix of the accelerated Bob (Rob) degenerates into the superposition of the right and left wedges of the Rindler spacetime ($I$ and $II$), and the resultant density matrix will be in the form of a three-qubit state $\rho_{(A,I,II)}$, where, A, represents Alice's party, while $I$ and $II$ represent the contribution from the right and the left wedge of the Rindler spacetime. Since we have no access to the region $II$, we trace out this region and end up with $\rho_{(A,I)}$. Note that in the case of black holes, the region $II$ belongs to the inside of the even horizon of black holes, which we have no access to. Nevertheless, considering the acceleration frame, we find the density matrix of Alice and Rob as
	
		\begin{equation}
\rho_{A,I}=\left(\begin{array}{cccc}
	\rho_{11} & 0 & 0 & \rho_{14} \\
	0 & \rho_{22} & 0 & 0 \\
	0 & 0 & \rho_{33} & 0 \\
	\rho_{14}^{*} & 0 & 0 & \rho_{44}
	\end{array}\right).
	\label{Werner-Unruh}
	\end{equation}
	where 
		\begin{eqnarray} \notag
	\rho_{11}=\rho_{22}= \frac{1+p}{2}\cos^2(r), \\ \notag
	\rho_{33}=\frac{1-p}{2}+\frac{1+p}{2}\sin^2(r), \\ \notag
	\rho_{44}= \frac{1-p}{2}+\frac{1+p}{2}\sin^2(r), \\ \notag
	\rho_{14}=\rho_{14}^{*}=p \cos(r).
\end{eqnarray}

	\subsection{Quantum entanglement}

	\begin{figure}
		\centering
		\includegraphics[width=3 in]{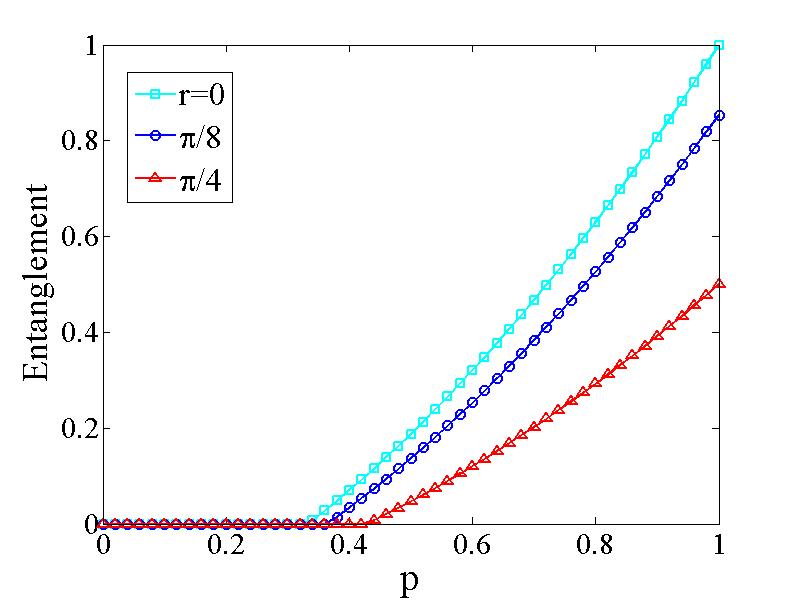}
		\caption{Concurrence versus $p$ for various $r$. The red, blue and green curves correspond to $r=0$,  $r=\pi/8$ and  $r=\pi/4$, respectively.
		}
		\label{fig:2}
	\end{figure}

To have a better insight into the quantum correlations and consider the entanglement of the system we quantify the entanglement. One can use concurrence as a measure of entanglement in a bipartite system of two-qubits. The entanglement of a two-qubit system can be quantified by the concurrence \cite{wootters1998entanglement}
	\begin{align}
	C=\textrm{max}\{0,\lambda_1-\lambda_2-\lambda_3-\lambda_4\},
	\end{align}
	where $\lambda_i$ are the eigenvalues of the Hermitian matrix
	$$
	R=\sqrt{\sqrt{\rho}\tilde{\rho}\sqrt{\rho}}.
	$$
	The operator $\tilde{\rho}$ is defined as 
	$$
	\tilde{\rho}=(\sigma_y\otimes\sigma_y)\rho^*(\sigma_y\otimes\sigma_y).
	$$
	Here, $\rho^*$ is the complex conjugate
	of $\rho$, and $\sigma_y$ is the spin-flip operator ($y$ component of the Pauli matrices).
	
	In the definition of the concurrence $C$, the eigenvalues are given in non-increasing order, expressed as 
	$$
	\lambda_1\geq\lambda_2\geq\lambda_3\geq\lambda_4.
	$$
The entanglement measure, concurrence, ranges between 0 and 1. For a maximally entangled state, we have $C=1$ and for a separable state $C=0$.

	Using concurrence to quantify entanglement between Alice and Rob, we find its explicit form for the state (\ref{Werner-Unruh}) to be
	
	\begin{align}\label{conc}
	C= \frac{ \cos^2(r)}{4}\textrm{max}    
	\Bigg\{0,4p^2+2p-2+(1-p^2)\cos^2(r) \Bigg\}.
	\end{align}
	Therefore, we find from this relation 
	the criterion of separability of the state as 
		\begin{equation}
	p\leq\frac{2-\cos^2r }{4-\cos^2r}.
	\end{equation}
	This also agrees with the separability 
	criterion that we can attain from
	Positive Partial Transpose (PPT) investigations (see Appendix for more details) \cite{horodecki2001separability, Peres_1996}. As readily can be seen, for $r=0$, this criterion reduces to $p\leq 1/3$, which agrees with the analyses of 
	Werner state for non-accelerated setting \cite{werner1989quantum,miranowicz2004violation}. As the acceleration increases, the separability range of the parameter $p$ increases, and its maximum can be attained for $r=\pi/4$ as $p\leq 3/7$.
	The behavior of the concurrence versus $p$ is presented in Fig. \ref{fig:2}.
	Accordingly, the Unruh acceleration decreases the entanglement in the system.
	An an specific setting, for $p=1$, concurrence reduces to $C=\cos^2r$ which gives $C=1/2$, for $r=\pi/4$, as is shown in Fig. \ref{fig:2}.

	\subsection{Violation of Bell inequality}

	Bell inequalities can discriminate between the local hidden variable (LHV) description of quantum mechanics and the one with no LHV theory. The most commonly used Bell-inequality is the so-called CHSH inequality. Defining the Bell (CHSH) operator as \cite{clauser1969proposed,horodecki1995violating}
	$$
	\hat{B}=A_{1} \otimes\left(B_{1}+B_{2}\right)+A_{2} \otimes\left(B_{1}-B_{2}\right),
	$$
	where Alice's (Bob's) observables are denoted as $A_{1}$ or $A_{2}\left(B_{1}\right.$ or $\left.B_{2}\right)$ with eigenvalues $\pm 1.$ 
	Given the above relation, the Bell-CHSH inequality reads \cite{clauser1969proposed,horodecki1995violating}
	$$
	|\left\langle B_{\mathrm{CHSH}}\right\rangle_{\rho}|=|\operatorname{tr}(\rho \hat{B})| \leq 2.
	$$
	Violation of this inequality, i.e., $|\left\langle B_{\mathrm{CHSH}}\right\rangle_{\rho}|> 2$, indicates Bell nonlocality in a quantum state. 
	The maximum violation of the Bell inequality, $B_{\max }(\rho)$, is given as  $B_{\max }(\rho)=$ $\max |\left\langle B_{\mathrm{CHSH}}\right\rangle_{\rho}|$. This quantity, in turn, can be expressed in terms of $M(\rho)$ with $B_{\max }(\rho)=2 \sqrt{M(\rho)}$ \cite{clauser1969proposed,horodecki1995violating}.
	For the density matrix of the form
	
	\begin{equation}
	\rho^{}=\left(\begin{array}{cccc}
	\rho_{11} & 0 & 0 & \rho_{14} \\
	0 & \rho_{22} & \rho_{23} & 0 \\
	0 & \rho_{23}^{*} & \rho_{33} & 0 \\
	\rho_{14}^{*} & 0 & 0 & \rho_{44}
	\end{array}\right)
	\end{equation}
	the explicit form of $M(\rho)$ can be attained as 
	
	\begin{align}
	\nonumber
	M(\rho)=\max \{& 8(|\rho_{14}|^{2}+|\rho_{23}|^{2}) 
	\\
	&	,\left(\rho_{11}+\rho_{44}-\rho_{22}-\rho_{33}\right)^{2} 	+4\left(|\rho_{23}|+|\rho_{14}|\right)^{2}\}.
	\end{align}
	
Considering the density matrix (\ref{Werner-Unruh}), we obtain $	M(\rho_{A,I})$ as a function of $p, r$
	
	\begin{align}
	\nonumber
	M(\rho_{A,I})&=\text{max}\left\{ 2p^2\cos^2r , p^2\cos^4r+p^2\cos^2r\right\}
	\\
&=2p^2\cos^2r.
	\end{align}
	
	Given $r\in (0,\pi/4)$, we obtain the condition for the state being Bell non-local as
	
	\begin{equation}
	p>\frac{1}{\sqrt{2}\cos{r}}.
	\end{equation}

	\section{QSE of a two-qubit state}
	
	In this section, we review the concept of QSE and  maximal steered coherence (MSC). To this end, we start with the general form of the two-qubit state $\rho_{AB}$, represented in the Pauli basis. Such a density matrix can be expressed as
	\begin{align}\label{pauli-basis}
	\rho_{AB}=\frac{1}{4} \bigg[I\otimes I+\mathbf{a}.\boldsymbol{\sigma}\otimes I+I\otimes \mathbf{b}.\boldsymbol{\sigma}+\sum_{m,n=1}^{3}T_{nm}\sigma_{n}\otimes \sigma_{m}\bigg],
	\end{align}
	where $I$ is the identity operator, and $\sigma_{j}$, with $j=1,2,3$, is $j$th element of the Pauli matrices. $\boldsymbol{\sigma}$ represents the vector of these three Pauli matrices.
	Furthermore, $\mathbf{a}=\mathrm{tr}(\rho_{AB} \boldsymbol{\sigma}\otimes I)$ and $\mathbf{b}=\mathrm{tr}(\rho_{AB} I\otimes \boldsymbol{\sigma})$ provide the local Bloch vectors of the qubits.
	The bipartite correlations are determined by the matrix elements \cite{horodecki1996information}
	\begin{eqnarray}
	T_{nm}=\mathrm{tr}(\rho_{AB} \sigma_{n}\otimes \sigma_{m}).
	\end{eqnarray}
	
	If we perform a local measurement on Bob's qubit, Alice's state becomes steered. Hence, considering all possible local measurements by Bob, Alice's steering ellipsoid $\varepsilon_{A}$
	is centered at \cite{jevtic2014quantum}
	\begin{eqnarray}
	C_{A}=\frac{\mathbf{a}-T\mathbf{b}}{1-\mathbf{b}^2}.
	\end{eqnarray}
	Thus, the ellipsoid matrix is given by
	\begin{eqnarray}
	Q_{A}=\frac{(T-\mathbf{a}\mathbf{b}^{T})}{1-\mathbf{b}^2}(1+\frac{\mathbf{b}\mathbf{b}^{T}}{1-\mathbf{b}^2})(T^{T}-\mathbf{b}\mathbf{a}^{T}),
	\end{eqnarray}
	where the eigenvalues of $Q_{A}$ are the squares of the ellipsoid semiaxes $s_{i}$ and its eigenvectors provide the orientation of these axes.
	
	Alternatively, when Bob is steered by Alice's local measurements, we can obtain Bob's steering ellipsoid $\varepsilon_{B}$
	by exchanging $A$ and $B$. Thus, his QSE ($\varepsilon_{B}$) is centred at
	\begin{eqnarray}
	C_{B}=\frac{\mathbf{b}-T^{T}\mathbf{a}}{1-\mathbf{a}^2},
	\end{eqnarray}
	and his ellipsoid matrix is
	\begin{eqnarray}
	Q_{B}=\frac{(T^{T}-\mathbf{b}\mathbf{a}^{T})}{1-\mathbf{a}^2}(1+\frac{\mathbf{a}\mathbf{a}^{T}}{1-\mathbf{a}^2})(T-\mathbf{a}\mathbf{b}^{T}).
	\end{eqnarray}
	
	Now, we investigate the set of so-called canonical states, which have particular importance in the steering ellipsoid formalism \cite{jevtic2014quantum,milne2014jennings}.
	This canonical state, $\widetilde{\rho}_{AB}$, corresponds to a two-qubit state in which Alice's marginal is maximally mixed.
	We perform the stochastic local operations and classical communication (SLOCC) operator on qubit $A$ which transforms $\rho_{AB}$ to a canonical state
	\begin{align}
	\nonumber
	\rho_{AB}&\longrightarrow \widetilde{\rho}_{AB}
	=\big(\frac{1}{\sqrt{2\rho_{A}}}\otimes I \big)\rho\big(\frac{1}{\sqrt{2\rho_{A}}}\otimes I \big)^{\dag}
	\\
	&=\frac{1}{4} \bigg(I\otimes I+I\otimes \widetilde{\mathbf{b}}.\boldsymbol{\sigma}+\sum_{m,n=1}^{3}\widetilde{\mathbf{T}}_{nm}\sigma\otimes \sigma\bigg).
	\end{align}
	
	Since SLOCC operations on Alice do not affect Bob's steering ellipsoid, the same $\varepsilon_{B}$ can describe the characteristics of both $\rho_{AB}$ and the canonical state $\widetilde{\rho}_{AB}$.

	Now, let us consider a bipartite quantum state $\rho_{AB}$, such that the eigenstates of the reduced density matrix $\rho_{B}$ are denoted as $\Xi=\{|\chi_{i}\rangle\}$.
	When we perform a positive operator-valued measure (POVM) on Alice and obtain an outcome $M$, Bob's state is steered to $\rho^{M}_{B}:=\mathrm{tr}_{A}(M\otimes I \rho)/p_{M}$ with $p_{M}:=\mathrm{tr}(M\otimes I \rho)$, where $I$ represents the single qubit identity operator. The quantum coherence of the steered states $\rho^{M}_{B}$, as the summation of the absolute values of off-diagonal elements in the basis $\{|\chi_{i}\rangle\}$,  gives \cite{baumgratz2014quantifying}
	\begin{eqnarray}
	C(\rho^{M}_{B},|\chi_{i}\rangle)=\frac{1}{p_{M}}\sum_{i \neq j}|\langle \chi_{i}|\mathrm{tr}_{A}(M\otimes I \rho)|\chi_{j}\rangle|.
	\end{eqnarray}
	By considering all possible POVM operators on Alice, the set of all $\rho^{M}_{B}$ provides $\varepsilon_{B}$.
	Maximizing the coherence $C(\rho^{M}_{B},|\chi_{i}\rangle)$ over all possible POVM operators $M$ gives MSC as \cite{hu2016quantum}
	\begin{align}
	MSC(\rho^{M}_{B}):=\max_{M \in POVM} \left[\frac{1}{p_{M}}\sum_{i \neq j}|\langle \chi_{i}|\mathrm{tr}_{A}(M\otimes I \rho)|\chi_{j}\rangle|\right].
	\end{align}
	
	If $\rho^{M}_{B}$ is degenerate, $\Xi$ is not uniquely defined; however, MSC is defined over all possible POVM operators and taking  infimum over all possible reference basis $\Xi$ as \cite{hu2016quantum}
	\begin{align}
	\nonumber
	&MSC(\rho^{M}_{B}):=
	\\
	&\inf_{\Xi} \left\{ \max_{M \in POVM} \left[\frac{1}{p_{M}}\sum_{i \neq j}|\langle \chi_{i}|\mathrm{tr}_{A}(M\otimes I \rho)|\chi_{j}\rangle|\right] \right\}.
	\end{align}
	
	According to Ref. \cite{hu2016quantum}, MSC is the maximal perpendicular distance between a point on the surface of $\varepsilon_{B}$ and the reference basis.
	Specifically, when the input state is an X-state and reference basis is along an axis of $\varepsilon_{B}$, it was found that MSC is the length of the longest of the other two semiaxes. However, when the reference basis does not lay along the axis of $\varepsilon_{B}$, the MSC can be expressed by the length of the longest semiaxes of Bob's steering ellipsoid.

	\section{Quantum steering of two-qubit states in accelerating reference frame}
	\subsection{QSE of Werner-like state}

	Now that we have the relativistic density matrix in hand, we consider the steering of one of the parties. More specifically, we consider the steering ellipsoids of one of the qubits once the other is measured. 
	According to the previous section, if Alice performs a measurement on her qubit, we shall find Bob's QSE to be centered at $C_{B}$. Considering the Werner state above, we find $C_{B}$ for the density matrix $\rho_{(A,I)}$ to be

	\begin{eqnarray}
	C_{B}=\Bigg(0, 0, \displaystyle \frac{p\sin^2(r)}{\sin^2(r)+1}\Bigg).
	\end{eqnarray}
	
	This, indeed, provides the center of the Bob's ellipsoid once the measurement is performed by Alice. In order to determine the complete ellipsoid and the steered coherence, we further need the  ellipsoid matrix of Bob's density operator, for which we find
	\begin{eqnarray}
	Q_{B}=\left(
	\begin{array}{ccc}
	\displaystyle \frac{p^2}{\sin^2(r)+1} & 0 & 0\\
	0 & \displaystyle \frac{p^2}{\sin^2(r)+1} & 0\\
	0 & 0 & \displaystyle \frac{p^2}{(\sin^2(r)+1)^2}\\
	\end{array}
	\right).
	\end{eqnarray}

	Hence, the lengths of the semiaxes in $x_{1}$, $x_{2}$ and $x_{3}$  directions, which we denote respectively as $s_1$, $s_2$ and $s_3$, are given by
	
	\begin{equation}
	\begin{array}{l}
	\displaystyle s_{1}=s_{2}=\frac{p}{\sqrt{\sin^2(r)+1}},\\\\
	\displaystyle \quad s_{3}=\frac{p}{\sin^2(r)+1},
	\end{array}
	\end{equation}
	with this, we can find Bob's QSE which is steered by Alice as
	\begin{eqnarray}
	\varepsilon_{B}=\left\{|\left(
	\begin{array}{ccc}
	C_{B}(1)\\
	C_{B}(2)\\
	C_{B}(3)\\
	\end{array}
	\right)+\left(
	\begin{array}{ccc}
	s_{1} x_{1}\\
	s_{2} x_{2}\\
	s_{3} x_{3}\\
	\end{array}
	\right) | x\leq1 \right\}.
	\end{eqnarray}
	
	Using the relation for the center of the ellipsoid we have
	\begin{eqnarray}
	\varepsilon_{B}=\left\{|\left(
	\begin{array}{ccc}
	0\\
	0\\
	\frac{p\sin^2(r)}{\sin^2(r)+1}\\
	\end{array}
	\right)+\left(
	\begin{array}{ccc}
	s_{1} x_{1}\\
	s_{2} x_{2}\\
	s_{3} x_{3}\\
	\end{array}
	\right) | x\leq1 \right\}.
	\end{eqnarray}
	This determines the geometry of entire states that Bob's system can be steered to by the measurement performed via Alice on her system.

	As we pointed out earlier, MSC is given as the length of the longest semiaxes of Bob's state $\rho^{M}_{B}$, which has been steered by Alice. Therefore the MSC in the above system could be determined through
	\begin{eqnarray}
	MSC(\rho^{M}_{B})=\max\{s_{1}, s_{2}, s_{3}\}.
	\end{eqnarray}
	
	Consequently, we find for the MSC of the two-qubit system in Eq. (\ref{Werner-Unruh}) 
	
	\begin{eqnarray}
	MSC(\rho^{M}_{B})=\frac{p}{\sqrt{\sin^2(r)+1}}.
	\end{eqnarray}
	
	Therefore 
	\begin{eqnarray}
	\sqrt{2/3}p \leq MSC(\rho^{M}_{B})\leq p.
	\end{eqnarray}

	\begin{figure}
		\centering
		\includegraphics[width=3 in]{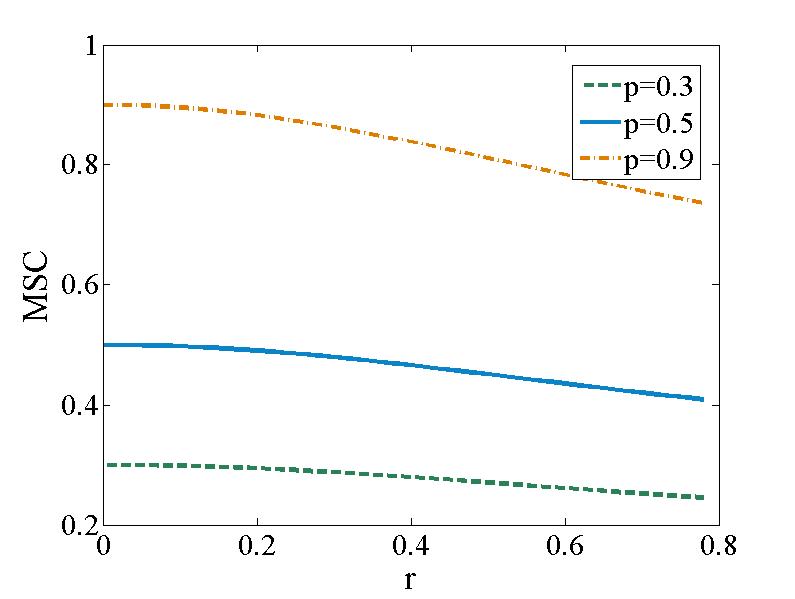}
		\caption{MSC versus $r$. The MSC starts from $p$ and monotonically decreases as the acceleration of the system get enhanced and when the acceleration goes to infinity the MSC asymptotically approaches $\sqrt{2/3}p$.
		}
		\label{fig3}
	\end{figure}

	\begin{figure*}
		\centering 
\qquad	 \textbf{(a)} 	\qquad 	\qquad \qquad	 \qquad	\qquad \qquad \qquad \qquad \textbf{(b)} 	\qquad \qquad		\qquad \qquad	 \qquad \qquad \qquad  \qquad \textbf{(c)} \qquad
		\\
		 {  
			\includegraphics[width=2.25in]{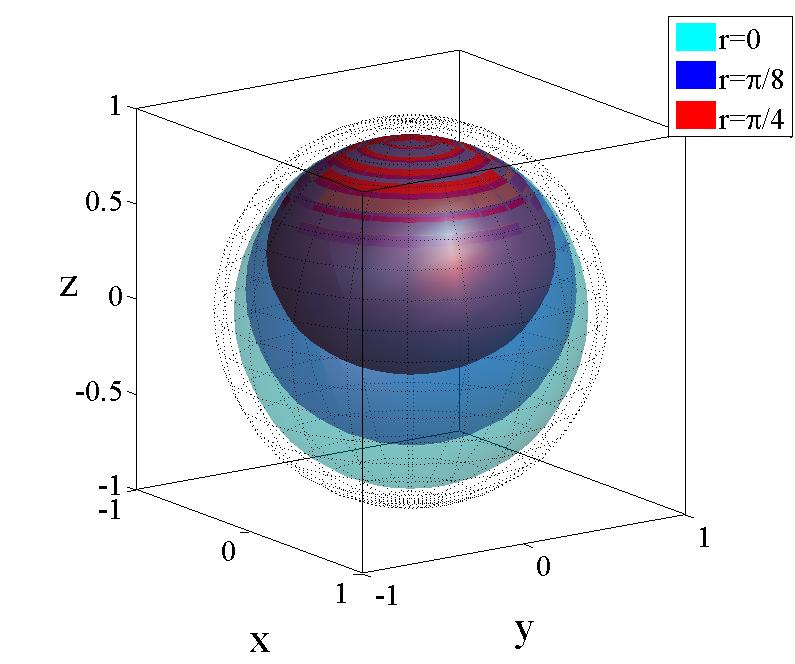}
			\label{fig:first_sub}
	}
	{ 
			\includegraphics[width=2.25in]{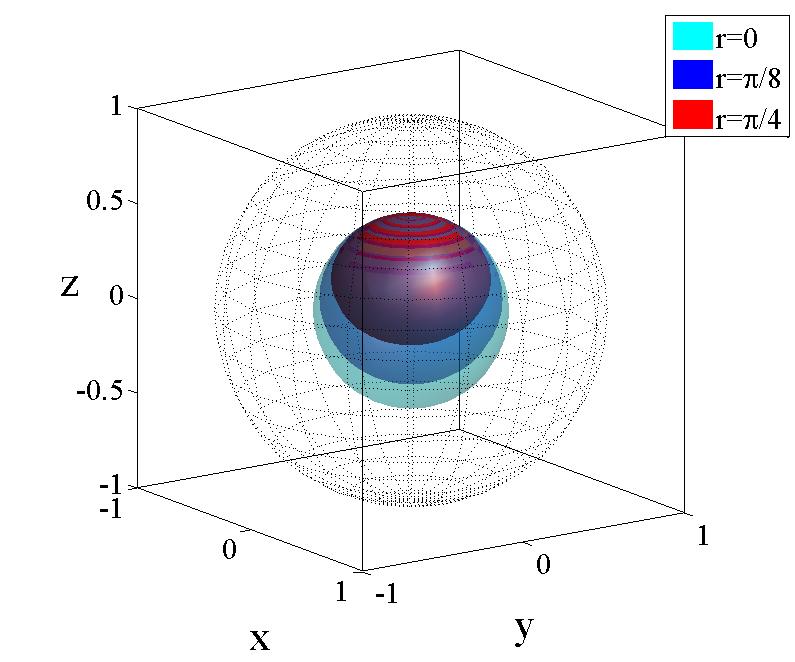}
			\label{fig:first_sub}
	}
	{ 
			\includegraphics[width=2.25in]{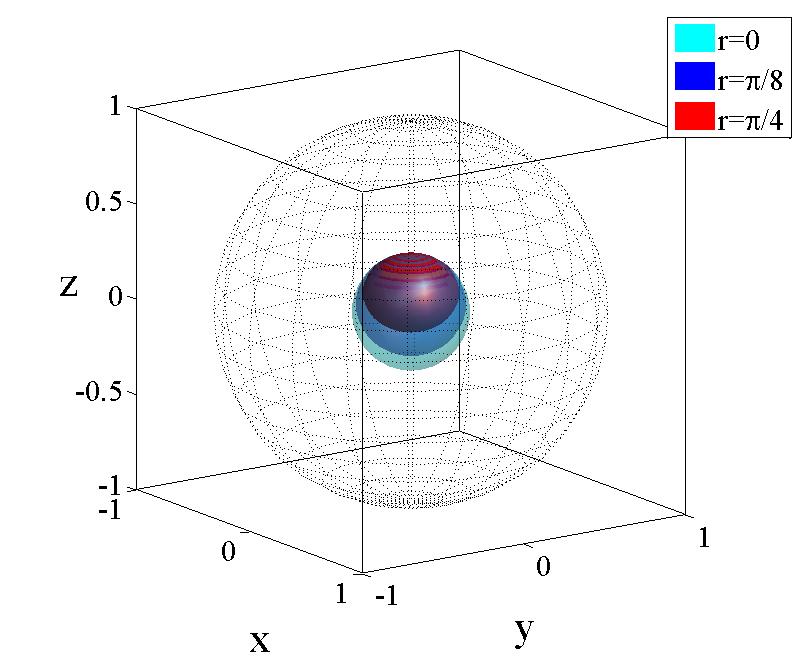}
			\label{fig:first_sub}
	}
		\caption{Bob’s quantum steering ellipsoid for various Unruh accelerations with (a) $p=0.9$, (b) $p=0.5$ and (c) $p=0.3$. The cyan, blue and red ellipsoid present QSE for $r=0, \pi/8$ and $\pi/4$, respectively. }
	\end{figure*}

	This relation provides an analytic result for the maximal steered coherence of Bob's state, and it demonstrates how exactly the steered coherence is affected by the acceleration of the subsystem. We plot the MSC (scaled by $p$) versus the acceleration parameter $r$ in Fig. (\ref{fig3}). As can be seen, MSC starts from $p$ and monotonically decreases by enhancement of the acceleration until it asymptotically approaches $\sqrt{2/3}p$ once the acceleration goes to infinity.

	With these analyses, we present the effect of acceleration on Bob's steered ellipsoid $\varepsilon_{B}$. We plot the ellipsoid for various states determined by $p$ in Fig. (4). We plot the ellipsoid $\varepsilon_{B}$ for the choices of the parameter $p=0.9,0.5,0.3$, and for $r=0,\pi/8,\pi/4$. We observe that the decrease in the parameter $p$ leads to the shrink of the ellipsoid. In other words, the size of the ellipsoids is related to the mixedness of the density matrix, such that it becomes smaller as the mixedness grows.
	On the other hand, the Unruh effect drastically affects the steering ellipsoids. As readily can be seen, by increasing $r$, the size of the ellipsoid shrinks.

	\subsection{Steerability of two-qubit Werner-like state}
	
We earlier mentioned that the Werner-state is entangled but not Bell nonlocal for $1/3<p<1/2$. Therefore, the critical value for the steerability of this state is determined by $p_{s}=1/2$. The steerability determines whether we can find a local hidden state (LHS) representation for the two-qubit system.
As was shown earlier, after transforming into the accelerated coordinate, the entanglement of the state (\ref{wl-rho}) decreases. Hence, we may expect that the critical value of steerability, $p_{s}(r)$, becomes larger than the original critical value $p_{s}(r=0)=1/2$.
 For a general 2-qubit state, the steerability can be determined using the "critical radius" method, which is developed by Nguyen, Nguyen, and Gühne \cite{NNG-2019}. 
 The basis of the idea is that if there exists a local hidden state representation for Bob, then the local hidden state must be in 2-dimensional Hilbert space and thus can be described by Bloch vector, and the distribution is on Bloch sphere. A crucial property is that the set of all states constructed from LHS model is convex. This enables finding a critical radius for the steerability of the density matrix determined by the Bloch sphere representation. 
 
 In general, by using convex optimization, the critical radius $r_c(\rho)$ of a state can be determined numerically with high precision. For some special cases, for example, Werner states, the critical radius can be determined analytically. The necessary and sufficient condition for a 2-qubit state $\rho$ to be steerable, i.e., to have a local hidden state (LHS) model, for the general projective measurements, is \cite{NNG-2019}
	\begin{equation}
	r_c(\rho)\geq 1.
	\end{equation}

	To determine the effect of the Unruh acceleration on the critical radius we need to decompose $\rho(A,I)$ in Pauli basis as represented by Eq. (\ref{pauli-basis}). To this end, we have  $\mathbf{a}^t=(0,0,0)$, $\mathbf{b}^t=(0,0,-\sin^2(r))$, and $T=\text{diag}\left(p\cos(r),-p\cos(r),p\cos^2 (r)\right)$. State of this type is called T-state, whose critical radius is shown to be \cite{NNG-2019}
	\begin{equation}\label{crit}
	r_c(\rho)=2\pi N_T |\text{det}(T)|, 
	\end{equation}
	where 
	$$
	N_T^{-1}=\int \frac{\vec{n}\cdot d\vec{S}}{(\vec{n}^tT^{-2}\vec{n})^2}.
	$$
	Here, $\vec{n}$ is a unit vector normal to Bloch sphere, and $N_T$ is a normalization factor for a uniform distribution on the Bloch sphere.
	Incorporating $T$ matrix in Eq. (\ref{crit}) we obtain the analytic form of the critical radius
	\begin{equation}
	r_c(\rho)=\frac{1}{p\left[\cos^2(r)+r\cot(r)\right]}.
	\end{equation}
	From $r_c(\rho)\geq 1$, the critical value for $p$ at fixed $r$ gives
	\begin{equation}
	p \leq p_s=\frac{1}{\left[\cos^2(r)+r\cot(r)\right]}.
	\end{equation}
	It can be verified that when $r=0$, $p_s$ degenerates to $1/2$ as is expected, and when $0<r<\pi/4$, $p_s(r)$ is always greater than $1/2$. In particular, for $r=\pi/4$, $p_s$ is $4/(2+\pi)\approx 0.778$. Therefore, enhancing the acceleration results in an increase in $p_s$.
	This demonstrates the impact of Unruh acceleration on steerability as a type of quantum correlation that cannot simply be characterized via entanglement measures.

	\section{Conclusion}

Quantum steering is a bizarre feature at the heart of quantum mechanics that provides important implications in understanding the nature of physical reality. On the other hand, the effect of the gravitational field on the quantum mechanical systems is a critical point in understanding the underlying foundations of physics. In this work, we considered the effects of Unruh acceleration on the quantum steering of the system. In particular, we studied the so-called quantum steering ellipsoid and the maximally-steered coherence in a non-inertial frame
and derived closed-form analytic expressions for the effect of the Unruh acceleration on these quantities. Moreover, we found the condition for the steerability of the system in this scenario.
Since the event horizon of the black hole, and generally any nonzero curvature, can be approximated by a locally accelerated reference frame, our study can shed new light on the quantum mechanical aspects of gravitational physics.

	\appendix*
	
	\section{PPT criterion}
	In this appendix, we present some details of the Positive Partial Transpose (PPT) criterion. To this aim, we consider a bipartite system, for which the density matrix is 
	
	$$\rho=\sum_{ij,kl} \rho_{ij,kl}\ket{i}_A\ket{j}_B \bra{k}_A\bra{l}_B.
	$$
	Then the partial transpose with respect to Bob's particle means transposing only Bob's basis, i.e.,
	\begin{equation}
	\rho^{\text{PPT}}=\sum_{ij,kl} \rho_{ij,kl}\ket{i}_A\ket{l}_B \bra{k}_A\bra{j}_B.
	\end{equation}
	Similarly, it can also be applied to Alice's particle. The PPT criterion states that if the state has no entanglement, then $\rho^{\text{PPT}}\geq0$, i. e., it has no negative eigenvalue. This applies to arbitrary bipartite states. Furthermore, for a 2-qubit state or qubit-qutrit state, this condition is sufficient for the system to be entangled.

\bibliographystyle{apsrev}
\bibliography{library}

\end{document}